\documentclass{elsart}
\usepackage{natbib}
\usepackage[T1]{fontenc}
\usepackage[latin1]{inputenc}
\usepackage{graphicx}

\makeatletter

\usepackage{amssymb}

\usepackage[english]{babel}
\makeatother
\begin{document}

\begin{frontmatter}

\title{Axial invariance of rapidly varying
diffusionless motions in the Earth's core interior}

\author{Dominique Jault}
\ead{Dominique.Jault@obs.ujf-grenoble.fr}
\address{LGIT, CNRS, Universit\'e Joseph-Fourier, BP 53, 38041 Grenoble
Cedex 9, France}

\begin{abstract}
Geostrophic jets propagating as Alfv\'en waves 
are shown to arise in
a rapidly rotating spherical shell permeated by a magnetic field
among the transient motions set up
by an impulsive rotation of the inner core. These axially invariant
motions evolve on a time-scale which is short compared to the magnetic
diffusion time. 
The numerical study is taken as illustrative of a more general point:
on such a fast time-scale the dimensionless number
appropriate to compare the rotation and magnetic forces is independent
of the magnetic diffusivity in contrast with the often used Elsasser number.
Extension of the analysis to non-axisymmetrical motions is 
supported by published
studies of dynamo models and magnetic instabilities.
\end{abstract}


\end{frontmatter}

\section{Introduction}
For the last ten years,
numerical simulations of the dynamo process in the Earth's core
have much changed the views on the interplay
between magnetic and rotation forces.
In particular, columnar flows almost invariant in the direction parallel
to the rotation axis and localized outside the imaginary cylinder tangent
to the inner core 
have been found very often 
even though the Elsasser number $\Lambda$, classically used
to estimate the ratio of magnetic to rotation forces, is of order 1
or larger (see e.g. \cite{olson99b,grote00}).
Alignment parallel to the rotation axis is caused by
the predominance of rotation forces.
Accordingly, columnar flows had been contemplated 
previously in the context of
weak-field models ($\Lambda \ll 1$) alone \citep{busse75}.
The ``strong field regime'' ($\Lambda = O(1)$) was illustrated by 
mean-field dynamo solutions, in which
the azimutal angular velocity showed instead large shears in the
direction parallel to the rotation axis \citep{brag78,hollerbach93,jault95}.
These early solutions were either steady or slowly varying
on the magnetic diffusion time in sharp contrast
with the current generation of dynamo solutions.

With large magnetic fields, as measured by $\Lambda$, only
the geostrophic part of the velocity field, symmetric
about the rotation axis, was expected to
be invariant in the direction parallel to the rotation axis.
\cite{brag70} singled out these motions in the context of magnetostrophic 
equilibrium, characterized
by the insignificance of inertial and viscous forces compared to magnetic,
rotation and pressure forces. 
He found that as these azimutal velocities
shear the magnetic field, they
are subject to a restoring force, provided by the magnetic field,
that ensures wave propagation.
This is the mechanism of
Alfv\'en waves and indeed geostrophic velocities in a rotating spherical
shell permeated by a magnetic field obey an equation of Alfv\'en
wave type save for geometrical factors.
\cite{brag70} assigned these torsional Alfv\'en waves to perturbations
with respect to a slowly evolving basic state characterized by the
cancellation of the total action of magnetic forces on the geostrophic
cylinders. That description sets the geostrophic velocities apart.
My aim, in this paper, is to defend another explanation for the
emergence of torsional Alfv\'en waves that can be generalized to
nonaxisymmetric motions, such as the almost
axially invariant vortices found in recent geodynamo solutions characterized
by strong but rapidly fluctuating magnetic fields. Other examples
are outlined in the discussion part.

In the next section, I introduce
the two dimensionless numbers 
$\Lambda$ and $\lambda$ that
measure
the relative strength of the magnetic and rotation forces, within
a rapidly rotating body permeated by a magnetic field.
I argue that on fast diffusionless time-scales, the appropriate
number is $\lambda$. 
This is illustrated in the third section, 
which constitutes the
main body of the article. The competition
between magnetic and rotation forces is studied in a rapidly
rotating spherical shell immersed in a magnetic
field. Specifically, the 
axisymmetrical transient
motions set-up  by an impulsive rotation of the inner core
are investigated for different values of $\lambda$ 
and the Elsasser number. 
This is followed by a general discussion, where
different problems are listed for which $\lambda$ rather
than $\Lambda$ is appropriate to compare magnetic and rotation forces.
The paper ends with concluding remarks.

\section{Lehnert versus Elsasser numbers}

\cite{elsasser46} argued that the magnetic field in the Earth's core
saturates when the magnetic force becomes comparable to the 
Coriolis force and suggested the characteristic strength

\begin{equation}
B= \left( \frac{2\Omega\rho}{\sigma}\right)^{1/2} \; ,
\label{B_Elsasser}
\end{equation}
where $\Omega$ is the angular velocity, $\rho$ is the density and
$\sigma$ is the electrical conductivity.
The Elsasser number, 
\begin{equation}
 \Lambda= \frac{\sigma B^2}{\Omega\rho}\; ,
\end{equation}
has subsequently been used to measure the relative strength of Coriolis and 
magnetic forces. In order to derive the relationship (\ref{B_Elsasser}),
the electrical current density $j$ is estimated as $\sigma U B$. 
This is obviously not valid when magnetic diffusion
is negligible compared to induction ($j\ll \sigma U B$).

Conversely, magnetic diffusion does not enter the physics of
plane magnetohydrodynamic waves, of length-scale $l$ that
\cite{lehnert54b} studied. He used
another dimensionless number $\chi_0=\lambda^{-1}$, with
\begin{equation}
 \lambda= \frac{B}{\Omega(\mu\rho)^{1/2}l}\; ,
\label{lambda}
\end{equation}
to measure the relative strength of magnetic and rotation forces. 
The parameter
$\lambda$, hereinafter referred to as the Lehnert number,
can be defined as the ratio, in a rapidly rotating 
and electrically conducting fluid
permeated by a magnetic field, of the period of the
inertial waves to the period of the Alfv\'en waves. Thus, the
typical frequency of diffusionless Alfv\'en waves is $\lambda \Omega$ and
the relationship
\begin{equation}
\lambda \ll 1 
\label{petit-lehnert}
\end{equation}
states that rotation forces dominate over magnetic forces on fast
time-scales.
\cite{cardin02} argued that both $\Lambda$ and $\lambda$ are
important to characterize geodynamo models.

In the spherical case, it is convenient to specify $\lambda$ using the 
outer radius $a$ as the length-scale $l$ in the definition (\ref{lambda}).
Denoting by $E_M$ the magnetic Ekman number $\eta / \Omega a^2$ 
and by $E$ the ordinary Ekman number
$\nu/\Omega a^2$,the relationship,
\begin{equation}
E_M +E \ll \lambda ,
\end{equation}
\label{atten}
\noindent
ensures that Alfv\'en waves are not rapidly damped by either magnetic
or viscous diffusion ($\nu$ and $\eta$ are respectively the viscous and
magnetic diffusivities).
Indeed, the ratio of
$\lambda$ to $E+E_M$ is the Lundquist number $S$ which indicates
how far Alfv\'en waves propagate before they are quenched by diffusion 
\citep{roberts67}. The two numbers $\lambda$ and
$\Lambda$ are related through the magnetic Ekman number:
\begin{equation}
\lambda ^2= \Lambda E_M .
\end{equation}

The value of $\lambda$ appropriate to the Earth's core is of the order of
$3. \times 10^{-5} - 2. \times 10^{-4}$. 
Indeed, $\lambda=10^{-4}$ corresponds to a magnetic field strength in the core
interior of the order of 3. mT. Usually quoted values range from
1. mT \citep{christensen06} to 4. mT \citep{starchenko02}. Using
$E_M = 4. \times 10^{-9}$ yields $S$  of the order of
$10^{4} - 5. \times 10^{4}$.

\section{Axisymmetric motions spawned in a spherical cavity by a
sudden impulse of the spin of the inner core}

The change of the relative strengths of the magnetic and rotation
forces according to frequency is well illustrated by the contrast
between transient and steady flows in a differentially rotating spherical
shell in the presence of a magnetic field. Static solutions have been published
in the case of dipolar magnetic field and small differential rotation, 
for which the structure of
the flow has been described according to the Elsasser number 
\citep{hollerbach94,dormy98}. \cite{kleeorin97} have theoretically 
investigated 
steady
linear solutions when the imposed magnetic field is potential 
and has dipole parity. 
They have identified several asymptotic regimes according to values 
of the Elsasser number in the small Ekman number limit.
Let us now study transient structures.

\begin{figure}
\centerline{
\includegraphics[clip=true,width=0.5\textwidth]{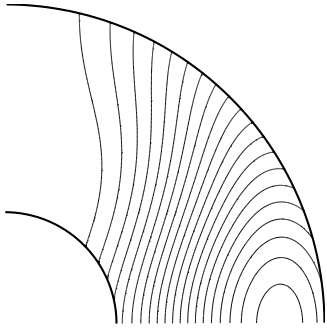}}
\caption{Field lines of the imposed poloidal axisymmetric field.}
\end{figure}

\subsection{Model and governing equations}

Consider an electrically conducting homogeneous fluid occupying
a spherical shell that
is immersed in an imposed steady magnetic field. 
The ratio of the inner shell radius $b$ to the outer radius $a$
is $b/a=0.35$ as in the Earth's core.
The solid inner core has the
same electrical conductivity as the fluid and the outer boundary is insulating.
The fluid is rotating with the constant angular velocity $\Omega$. 
The imposed magnetic field is chosen as:

\begin{eqnarray}
\mathbf{B} &= & B_0 \mathbf{\nabla} \times (A \mathbf{e_\phi}) \\
A &=& (j_1(\beta_{11} r)-0.3j_1(\beta_{12} r))P_1^1(\cos\theta)
 - 0.2\,j_3(\beta_{31} r)P_3^1(\cos\theta)
 \label{Bbasic}
\end{eqnarray}

\noindent
where $(r,\theta,\phi)$ are spherical coordinates, 
$P_l^1$ are Legendre functions,
$(j_l)_l$ is the set of
spherical
Bessel functions of the first kind and $\beta_{ln}$ is the {\it n}th root
of $j_{l-1}(\beta)=0$.
The basic field, which is shown in figure 1,
has been chosen with the aim of modelling torsional
oscillations in the Earth's fluid core. In this context, the important
quantity \citep{brag70} is
\begin{equation}
\{B_s^2\}(s)=\frac{1}{2\pi (z_T-z_B)}
\left(\oint \int _{z_B} ^{z_T} B_s^2 d z d\phi\right),
\end{equation}
\noindent
evaluated on geostrophic cylinders of radius $s$ and of top and bottom 
$z$-coordinates
respectively $z_T$ and $z_B$. Obviously, the choice (\ref{Bbasic})
is arbitrary. In contrast with the Earth's case and with the basic state
used by \cite{brag80} to model torsional oscillations, 
$\{B_s^2\}$ vanishes at $s=a$ because of the imposed
dipole symmetry with respect to the equatorial plane. 
Axisymmetry makes $\{B_s^2\}=0 $ at $s=0$ in contrast with the geophysical
case again.
In view of the
present study, the main characteristics of the field $\mathbf B$ 
defined by (\ref{Bbasic}) are that it is neither
parallel to the rotation axis nor rapidly decreasing with radius
as are current-free dipole fields.
The results presented below do
not depend on the details of the geometry of $\mathbf B$.

Study
the evolution of the velocity field $\mathbf u$ and of the magnetic field
deviation $\mathbf b$ after an impulsive increase of the angular rotation 
$\omega _b$ of the inner core, postulating symmetry about the axis of
rotation and dipole symmetry about the equatorial plane:
\begin{eqnarray}
u_r(r,\pi-\theta,\phi)&=&u_r(r,\theta,\phi), \; \;
u_\theta(r,\pi-\theta,\phi)=- u_\theta(r,\theta,\phi), \nonumber \\
u_\phi(r,\pi-\theta,\phi)&=& u_\phi(r,\theta,\phi), \; \;
b_r(r,\pi-\theta,\phi)=-b_r(r,\theta,\phi),  \\
b_\theta(r,\pi-\theta,\phi)&=& b_\theta(r,\theta,\phi),  \; \;
b_\phi(r,\pi-\theta,\phi)= -b_\phi(r,\theta,\phi). \nonumber
\end{eqnarray}
\noindent
The amplitude of the initial impulse
$\Omega _b$ is assumed to be
small enough so that the subsequent evolution of the dynamics
is independent of $\Omega _b$ within a scaling factor.
Using $B_0$ as unit of magnetic field, $a$ as length-scale and 
$a (\mu\rho)^{1/2}/B_0$ as unit of time,
the fields  $\mathbf u$ and $\mathbf b$ are
governed by the following linearised equations in the fluid region:

\begin{eqnarray}
\frac{\partial \mathbf{u}}{\partial t} 
+ 2 \lambda ^{-1} \mathbf{e_z} \times \mathbf{u} & = & - \nabla p 
+ 
(\mathbf{\nabla} \times \mathbf{B}) \times  \mathbf{b} +
 (\mathbf{\nabla} \times \mathbf{b}) \times  \mathbf{B} \nonumber \\
& & +
P_m \lambda\,\Lambda^{-1} \nabla ^2 \mathbf{u} \, , \;
\label{momentum}\\
\frac{\partial \mathbf{b}}{\partial t} & = & 
\mathbf{\nabla} \times (\mathbf{u} \times  \mathbf{B}) + 
\lambda\,\Lambda^{-1} \nabla ^2 \mathbf{b} ,
\label{induction}
\end{eqnarray}
where $P_m = \nu/\eta$ is the magnetic Prandtl number.
The field $\mathbf b$ is defined also in the inner solid region where:

\begin{equation}
\frac{\partial \mathbf{b}}{\partial t} = 
\lambda\,\Lambda^{-1} \nabla ^2 \mathbf{b} .
\end{equation}
Note that the steady-state solutions depend only on the two parameters
$\Lambda$ and $E$.
The velocity boundary conditions
\begin{eqnarray}
\mathbf{u} &=& 0\, , \; \; r=a \\
\mathbf{u} &=& s \, \omega _b (t) = s \,\Omega _b \, \delta (t-t_0), \; \; r=b
\end{eqnarray}
are written using the Dirac $\delta$ function and are 
appropriate to rigid boundaries.

The set of equations (\ref{momentum}) and (\ref{induction}) is discretized
and time-stepped from an initial state of rest:
\begin{equation}
\mathbf{u} = \mathbf{b} = \mathbf{0}
\end{equation}
\noindent
The Dirac $\delta$ function is approached as:
\begin{equation}
\frac{1}{\sqrt{\pi\epsilon}}e^{-(t-t_0)^2/\epsilon} .
\label{dirac}
\end{equation}
\noindent
It is a result of the simulations below that the solutions
are independent of the parameter $\epsilon$
provided that its value is set small enough.

A poloidal/toroidal decomposition
\begin{eqnarray}
\mathbf{u} &=& u_\phi \mathbf{e_\phi} +\mathbf{\nabla} \times
(u_p \mathbf{e_\phi}) \\
\mathbf{b} &=& b_\phi \mathbf{e_\phi} +\mathbf{\nabla} \times
(b_p \mathbf{e_\phi})
\end{eqnarray}
is employed. The variables are expanded in associated Legendre functions, i.e.
\begin{equation}
u_\phi(s,\theta)=\sum_{l=0}^{lmax}u_\phi^l(s) P_{2l+1}^1(\cos\theta)
\end{equation}
and then discretized in radius. The minimum truncation level
{\it lmax} is 120 whereas at least 450 unevenly spaced points are used
in the radial direction.

\begin{figure}
\centerline{
(\emph{a}) \includegraphics[clip=true,width=0.5\textwidth]{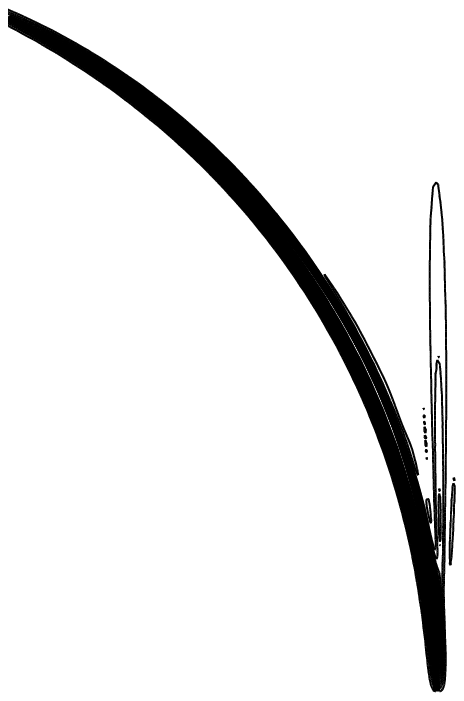}
(\emph{b}) \includegraphics[clip=true,width=0.5\textwidth]{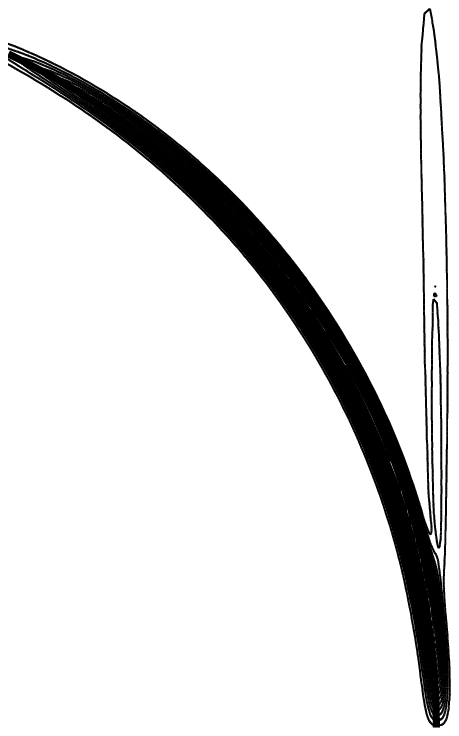}}
\centerline{
(\emph{c}) \includegraphics[clip=true,width=0.5\textwidth]{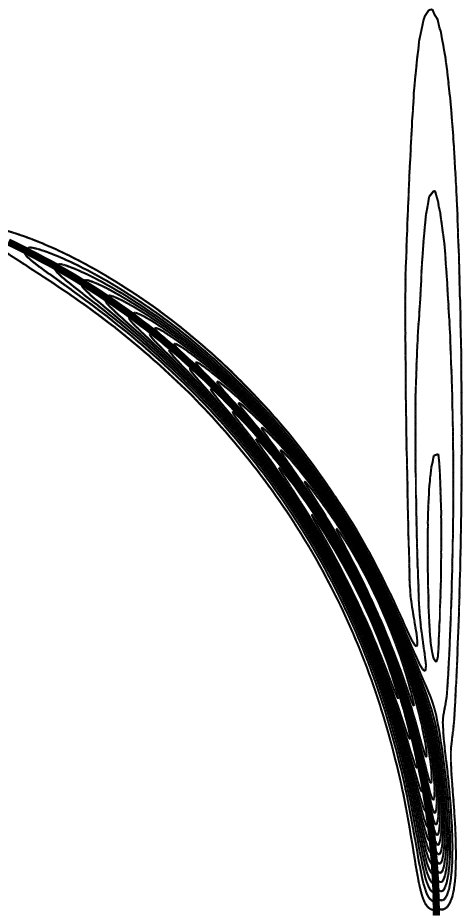}
(\emph{d}) \includegraphics[clip=true,width=0.5\textwidth]{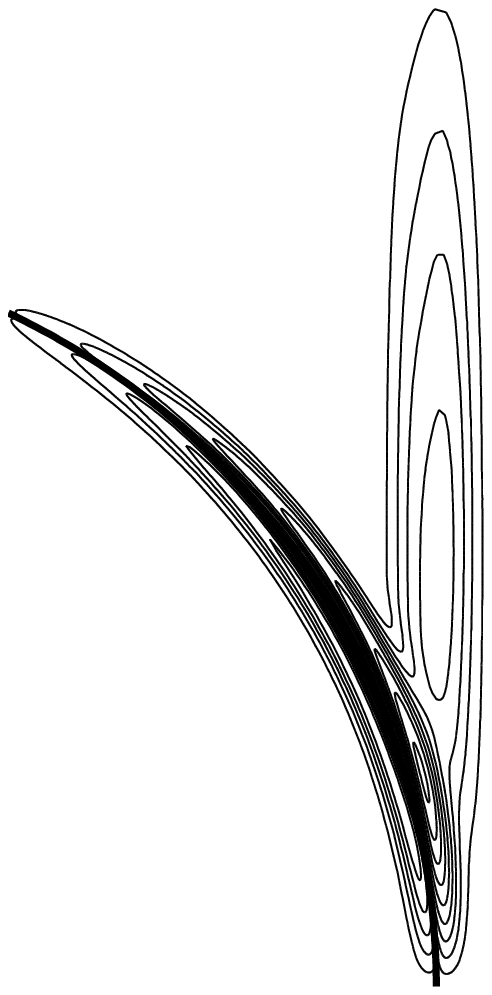}}
\caption{Contours of the induced azimutal magnetic field for 
$\lambda=1.72\times 10^{-4}$, $\Lambda=0.52$, 
$E=E_M=5.7\times 10^{-8}$ drawn at $t=8.6\times 10^{-3}$(a), 
$t=1.72 \times 10^{-2}$(b),
$t=3.44 \times 10^{-2}$(c) and $t=6.88 \times 10^{-2}$(d). The
contour intervals and the frame size 
are identical in all frames. The inner sphere
boundary is indicated with a thick line.}
\end{figure}

\begin{figure}
\centerline{
(\emph{a}) \includegraphics[clip=true,width=0.5\textwidth]{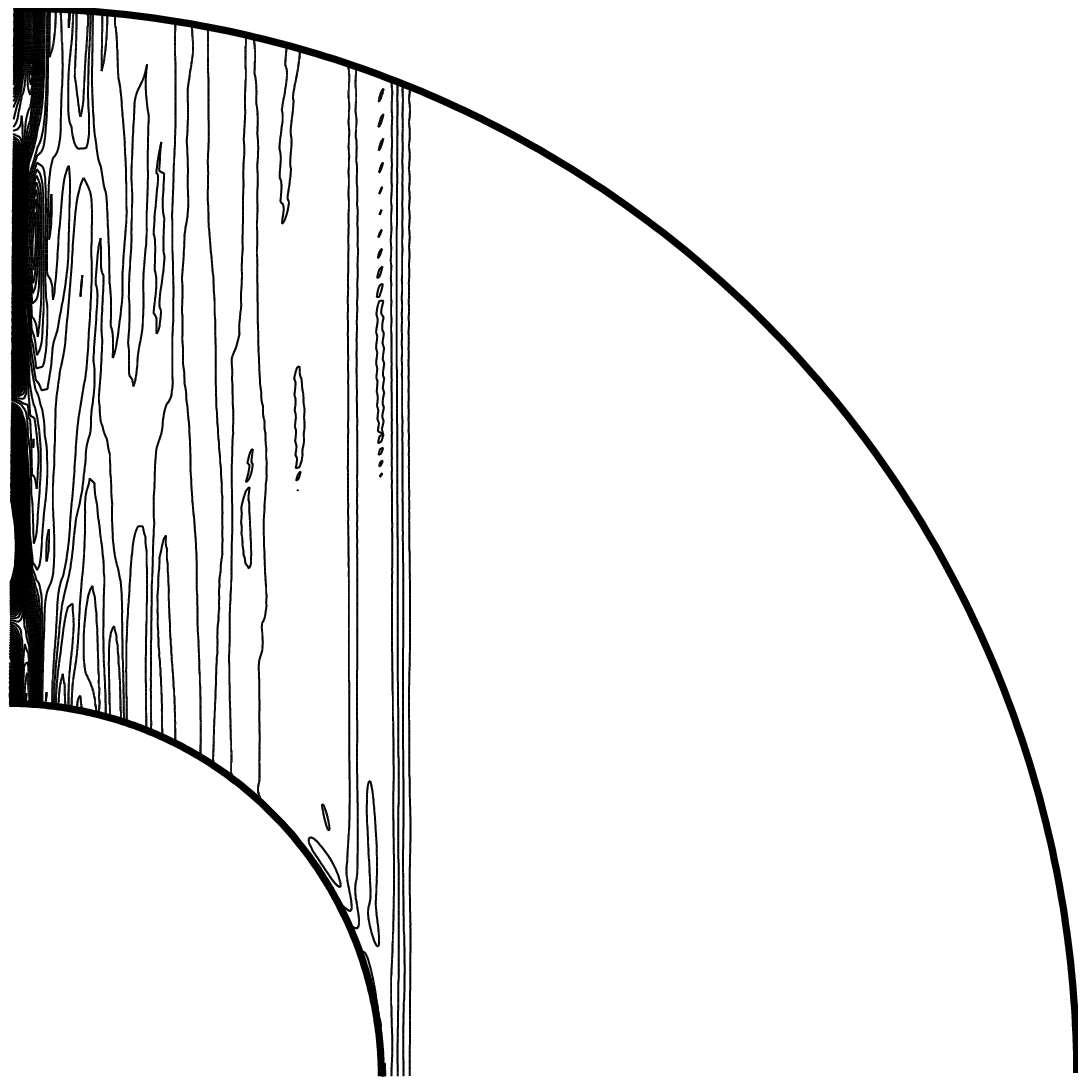}
(\emph{b}) \includegraphics[clip=true,width=0.5\textwidth]{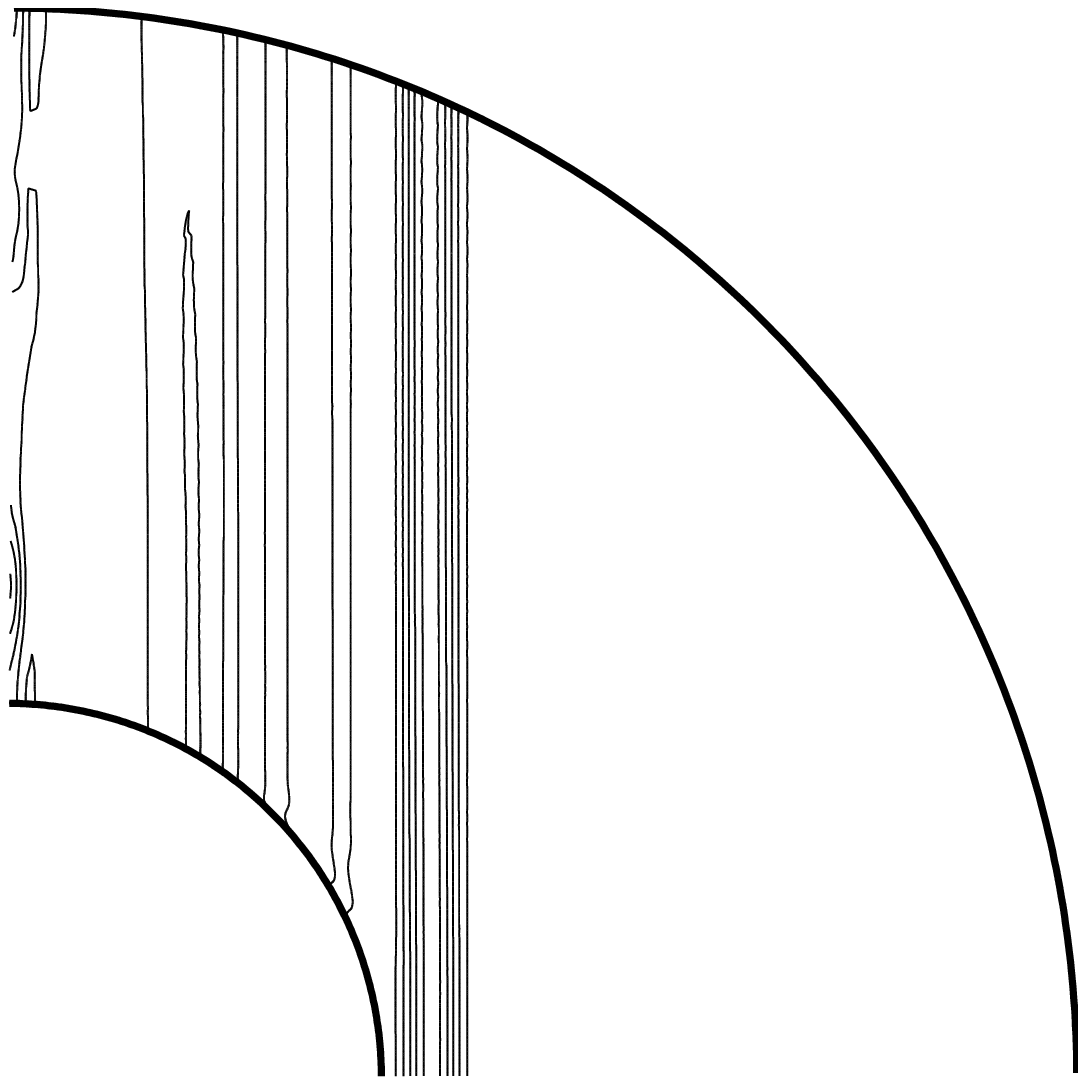}}
\centerline{
(\emph{c}) \includegraphics[clip=true,width=0.5\textwidth]{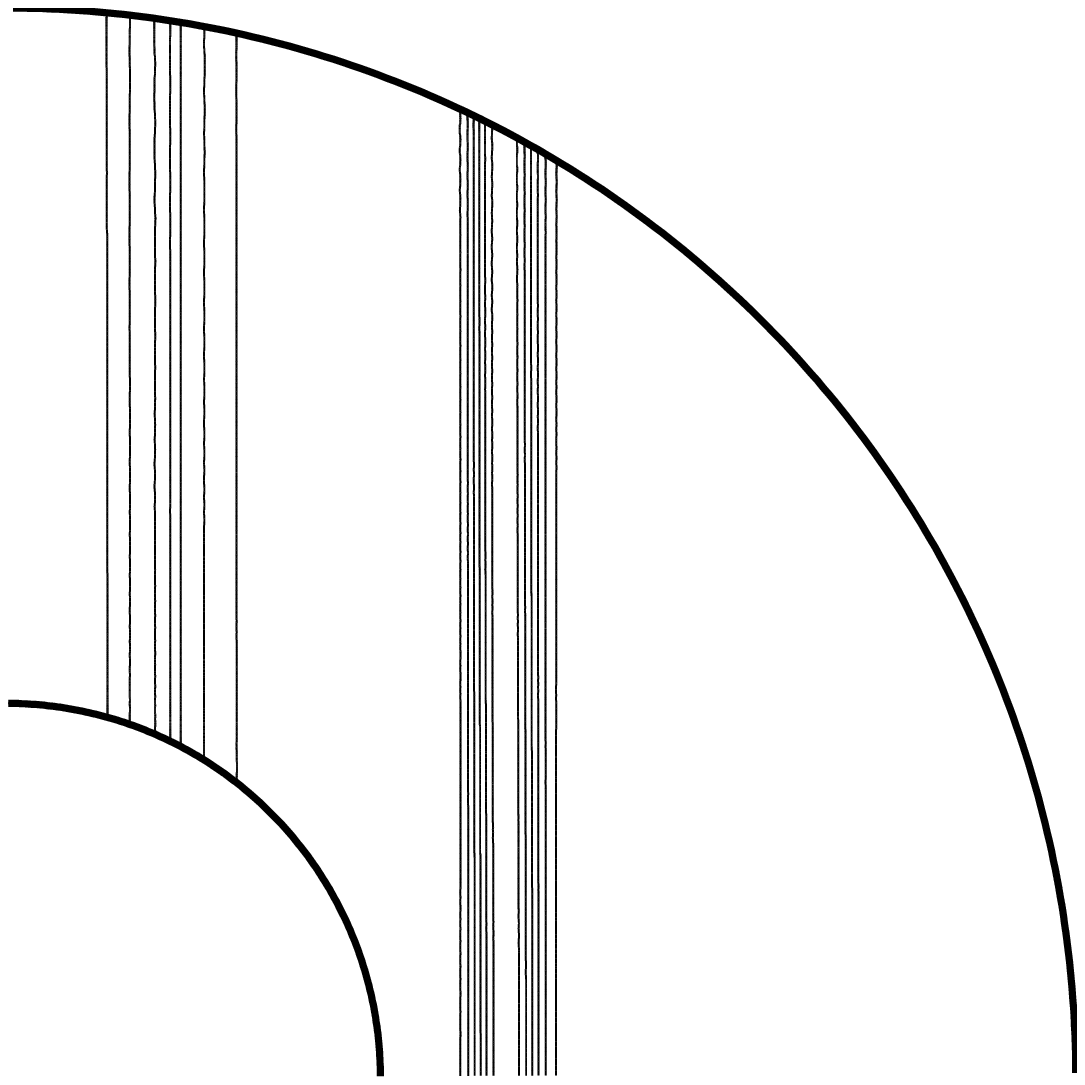}
(\emph{d}) \includegraphics[clip=true,width=0.5\textwidth]{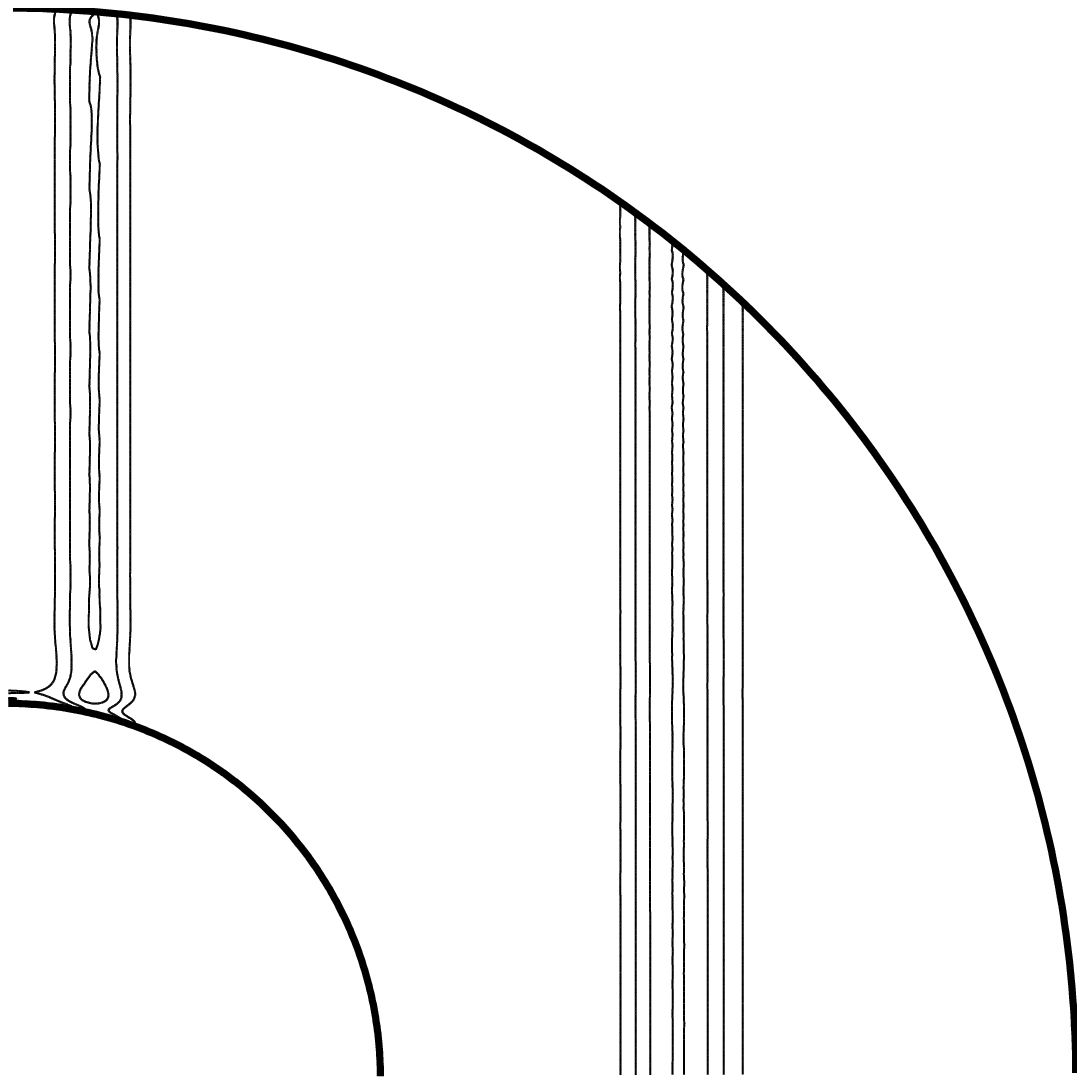}}
\caption{Contours of constant angular velocity for 
$\lambda=1.72\times 10^{-4}$, $\Lambda=0.52$, 
$E=E_M=5.7\times 10^{-8}$ drawn at $t=8.6\times 10^{-2}$(a), 
$t=0.26$(b),
$t=0.52$(c) and $t=1.03$(d). The
contour intervals are respectively $\Delta \omega$, $\Delta \omega/2$, 
$\Delta \omega/5$ and $\Delta \omega/10$ in the frames \emph{a}, \emph{b},
\emph{c} and \emph{d} in order to compensate for the attenuation of
the velocities.}
\end{figure}

\begin{figure}
\centerline{
(\emph{a}) \includegraphics[clip=true,width=0.5\textwidth]{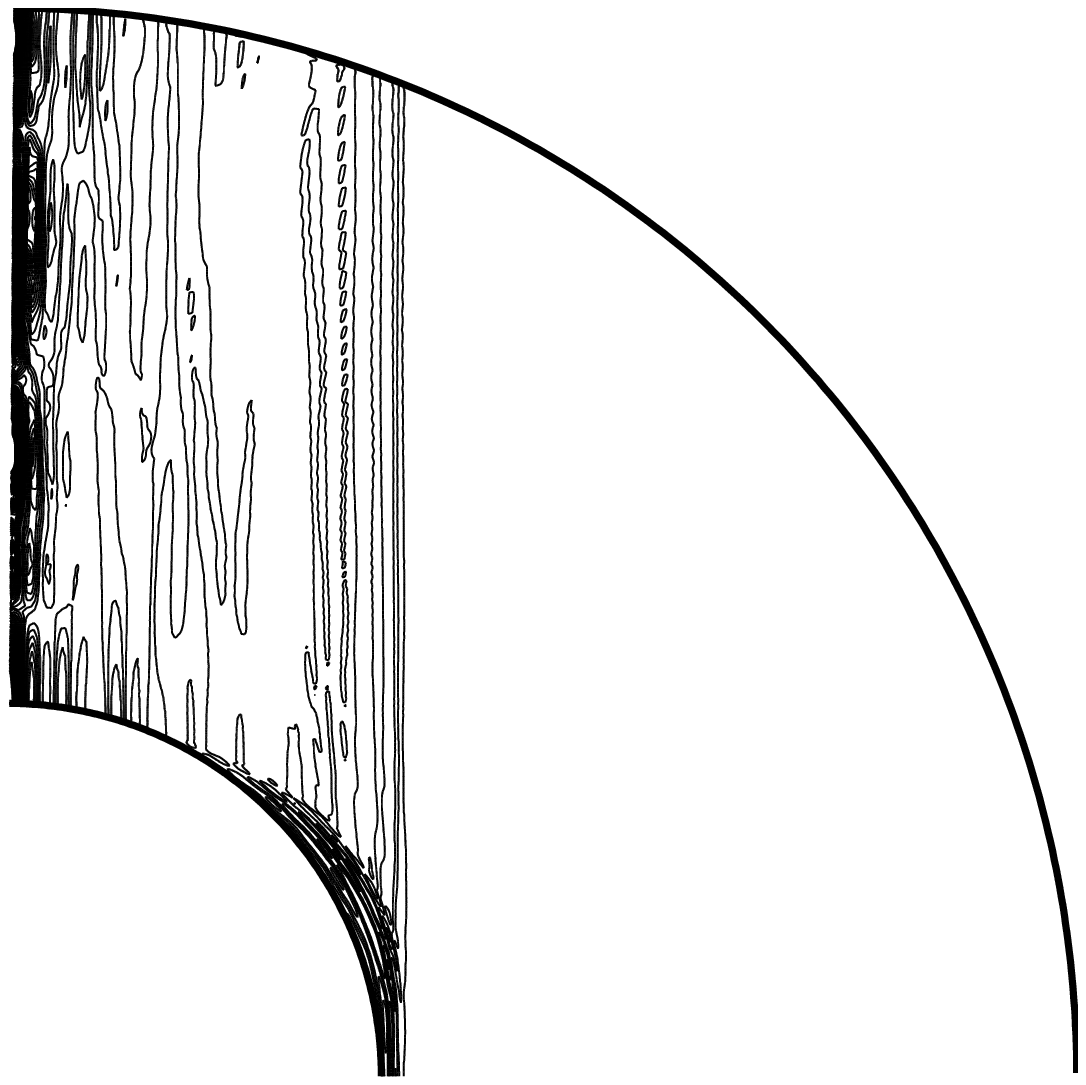}
(\emph{b}) \includegraphics[clip=true,width=0.5\textwidth]{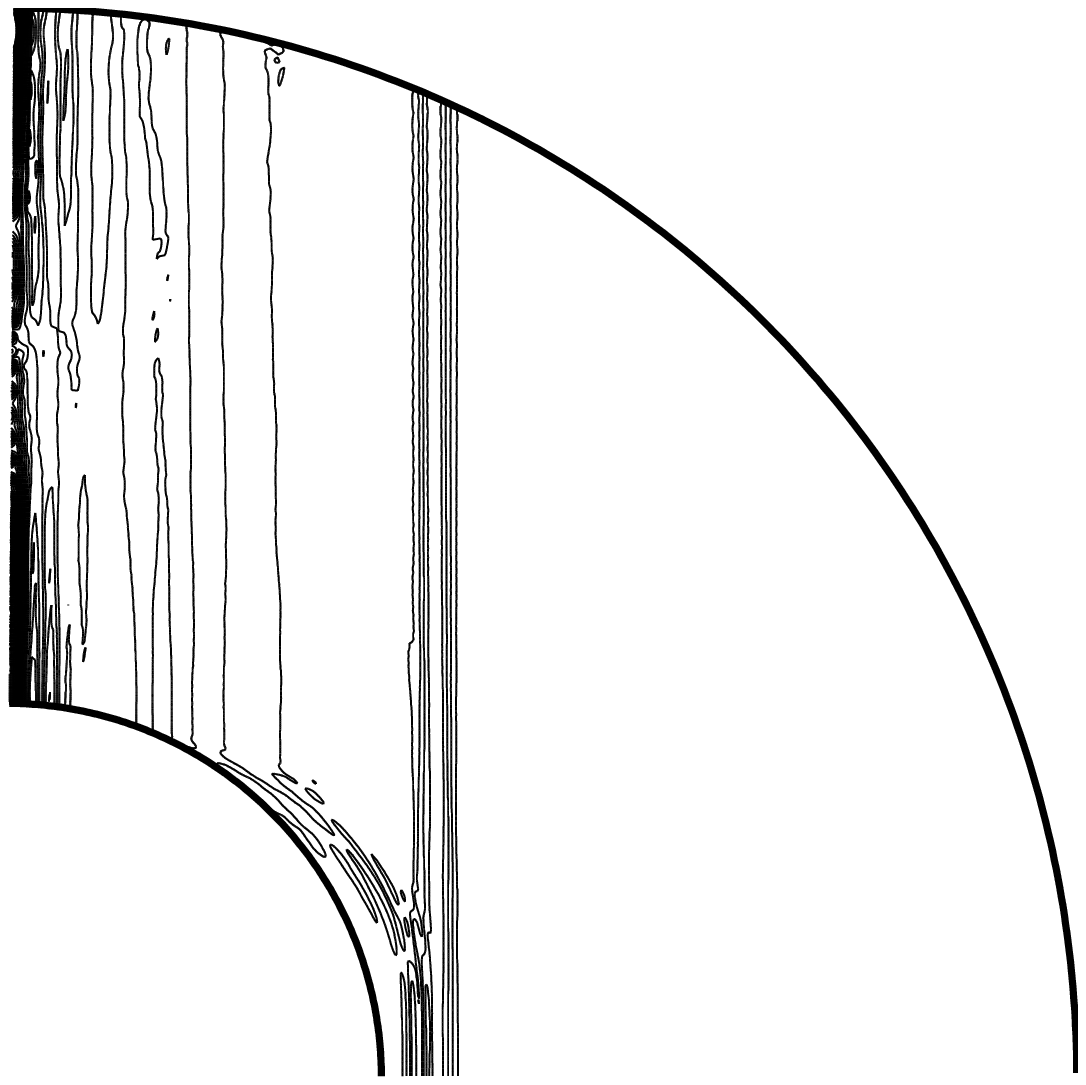}}
\centerline{
(\emph{c}) \includegraphics[clip=true,width=0.5\textwidth]{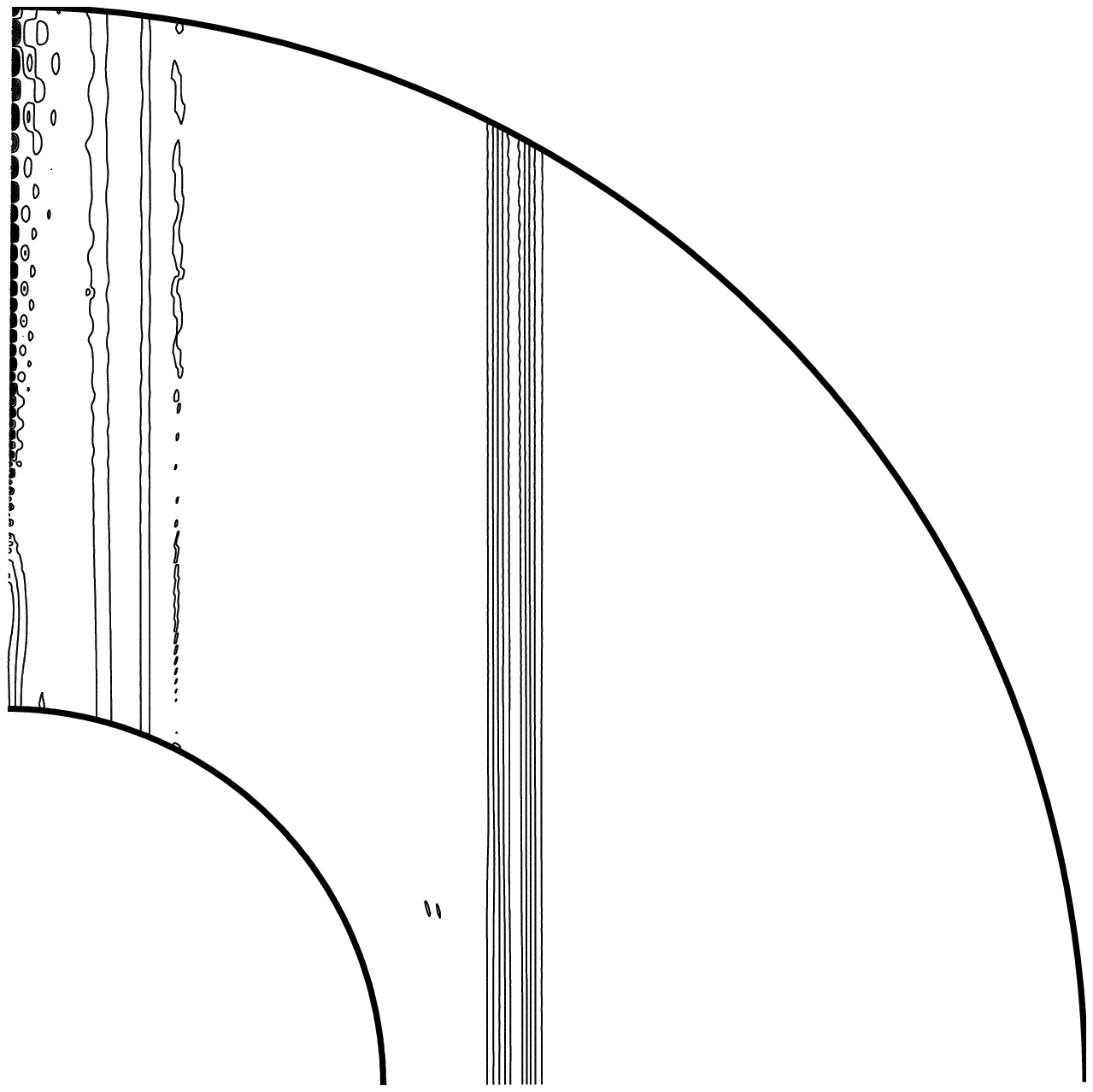}
(\emph{d}) \includegraphics[clip=true,width=0.5\textwidth]{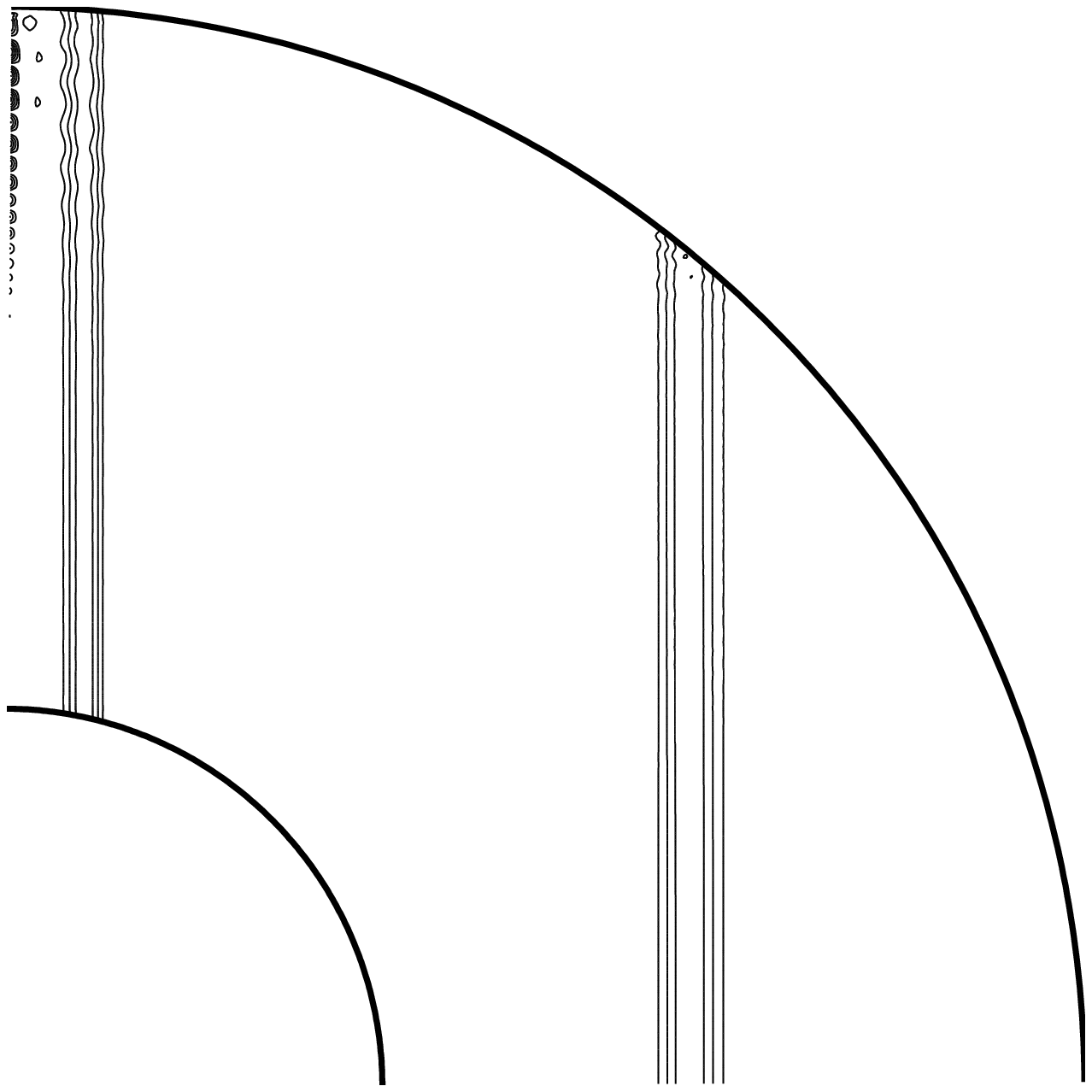}}
\caption{Same as figure 3 for
$\lambda=1.72\times 10^{-4}$, $\Lambda=6.5$, 
$E=1.42\times 10^{-8}$, $E_M=4.5\times 10^{-9}$ and same progression of
contour intervals.}
\end{figure}

\subsection{Formation and propagation of geostrophic jets}
\label{results}

Let us examine
a typical sequence of solutions for a small value of
$\lambda$. Following the initial impulse, an almost geostrophic shear
sets up, after a few revolutions,
at the cylindrical surface tangent to
the inner core, hereafter referred to as tangent cylinder. 
Induction of an azimutal magnetic field localized at the tangent cylinder
(compare the snapshot (a) to the snapshot (b) in figure 2
occurs about the end of this period during which the velocity field becomes
axially invariant. It starts from a source at the equator of the inner core.
The last two panels of figure 2 illustrates the following
period during which meridional electrical currents
parallel to the tangent cylinder intensify and loop further 
and further away from the tangent cylinder.
This is the most noticeable feature before 
the geostrophic shear splits up. 
The outer shear readily transforms into a jet propagating away from the tangent
cylinder towards larger cylindrical radii.
Thereafter, the velocity
within the jet becomes more and more invariant along $z$ as time elapses
(figure 3). 
On the other side of the tangent cylinder, a second shear propagates towards
the axis of rotation. Its propagation velocity slows down as $B_s$ decreases
to $0$ on the axis. In the event, the inner shear transforms also into a jet.
The comparison (figure 4) with a second sequence of
solutions for the same value of $\lambda$ but for $\Lambda$ multiplied by
a factor of 12.5 indicates that the dynamics outside the tangent cylinder is
almost independent of $\Lambda$. The flow remains geostrophic even though
$\Lambda$ is $O(1)$.
Note that steady flows do not reproduce this feature. Figure 5
shows zonal flows driven by rotating the inner sphere at a constant rate
for the two values of $\Lambda$ used to calculate the transient
solutions. For the largest value of $\Lambda$,
the angular velocity contours are not parallel to the rotation axis and
tend instead to follow the magnetic field lines, as prescribed by
Ferraro's law of isorotation. Steady solutions are established after
a period lasting a few time units $\lambda^{-1}\Omega^{-1}$ during which the
flow is geostrophic.

\begin{figure}
\centerline{
(\emph{a}) \includegraphics[clip=true,width=0.5\textwidth]{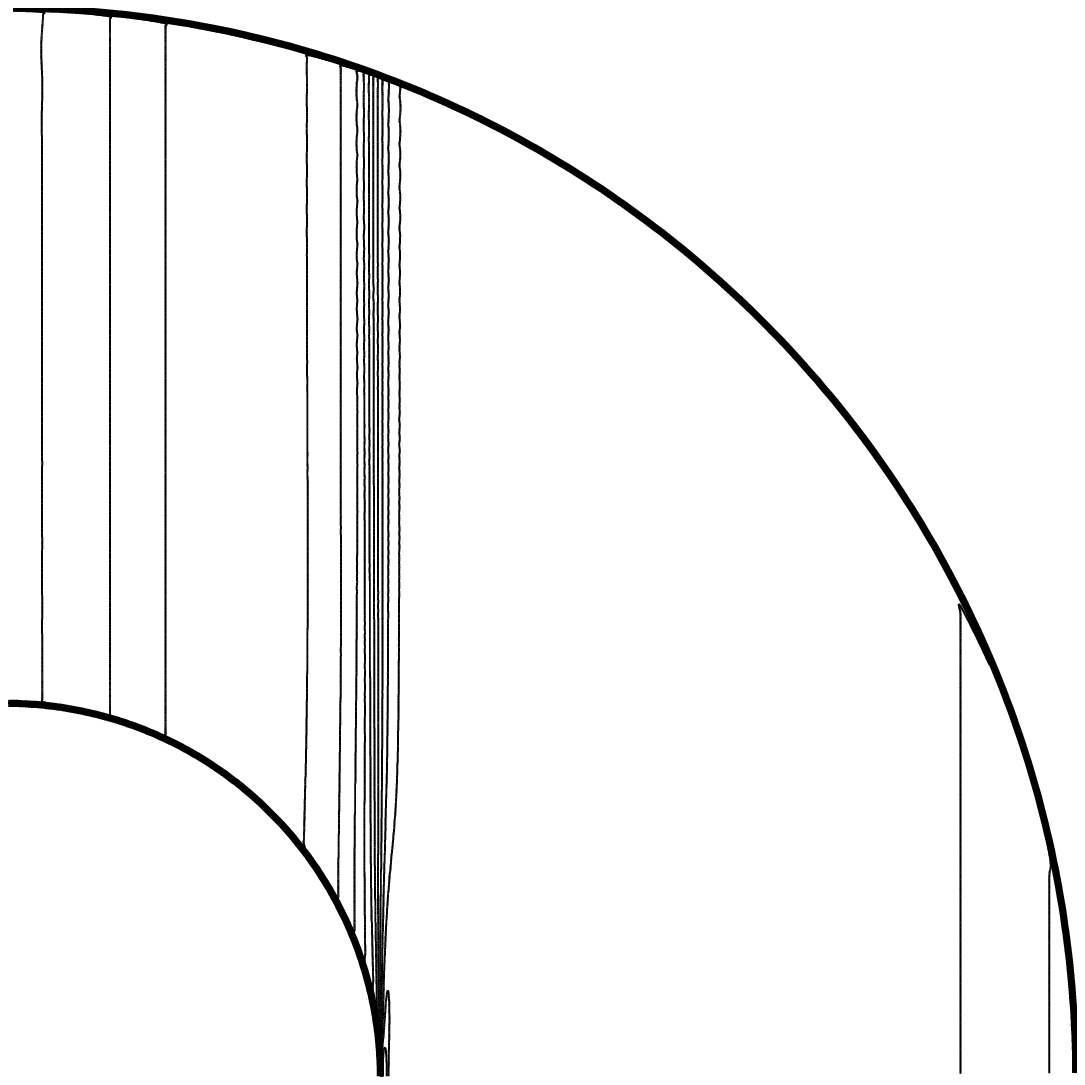}
(\emph{b}) \includegraphics[clip=true,width=0.5\textwidth]{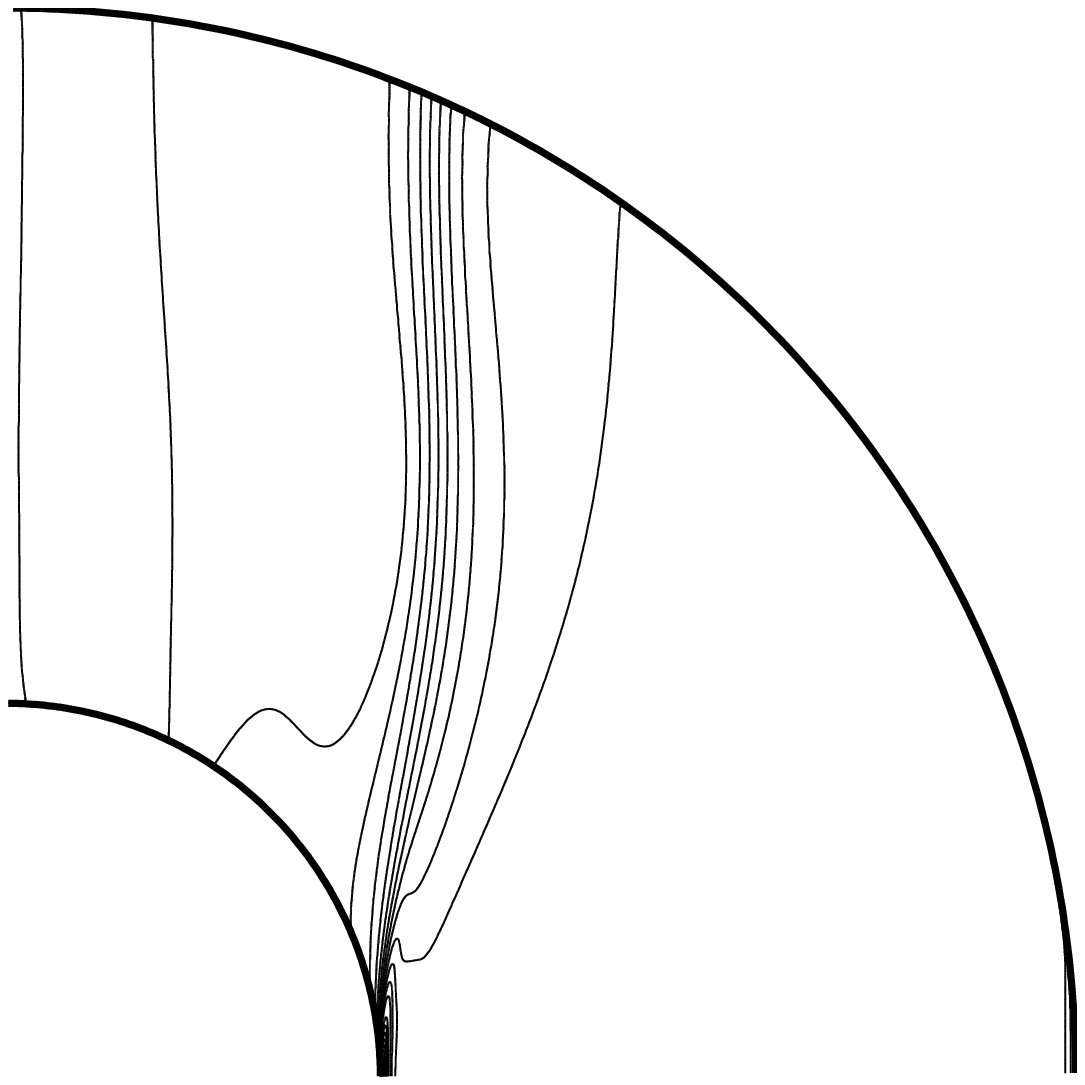}}
\caption{Steady azimuthal flow induced by rotating the inner and outer
boundaries at slightly different rates. Contours of constant angular
velocity for $E=4.5\times 10^{-7}$ and $\Lambda=0.52$ (a) $\Lambda=6.5$ (b).}
\end{figure}

\begin{figure}
\centerline{\hspace*{-0.05\textwidth}
\includegraphics[clip=true,width=0.8\textwidth]{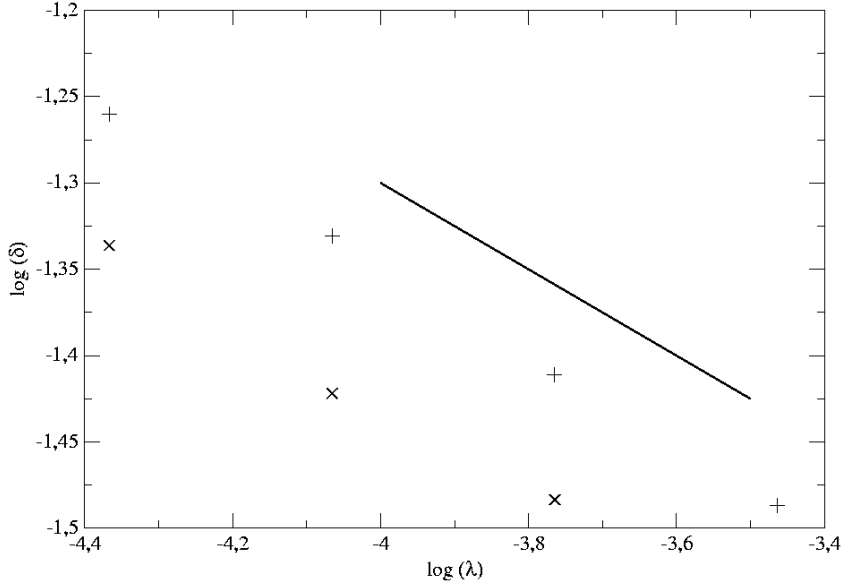}}
\caption{Scaling of the width $\delta$ of the outer geostrophic jet with
respect to the number $\lambda$. The magnetic Prandtl number is
$P_m=1$ and the Ekman number is $E=5.7\times 10^{-8}$ (crosses) or
$E=2.85\times 10^{-8}$ (+ signs). A line of slope $-1/4$ is shown
for comparison.}
\end{figure}

\begin{figure}
\centerline{\hspace*{-0.05\textwidth}
\includegraphics[clip=true,width=0.8\textwidth]{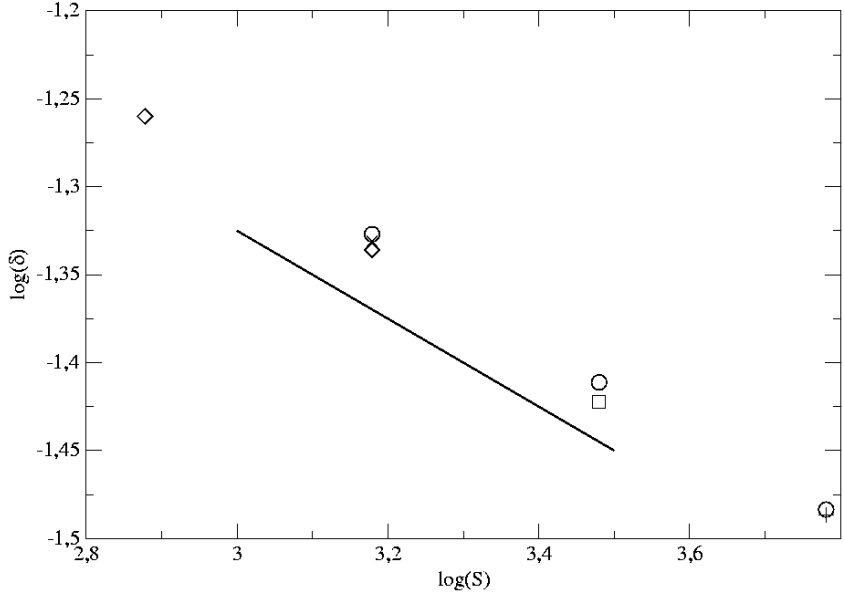}}
\caption{Scaling of the width $\delta$ of the outer geostrophic jet with
respect to the Lundquist number $S^\dag=\lambda/E_M$. 
The magnetic Prandtl number is
$P_m=1$ and $\lambda$ is $4.3 \times 10^{-5}$ ($\Diamond$), 
$8.6 \times 10^{-5}$ ($\times$), $1.72 \times 10^{-4}$ ($\circ$),
$3.44 \times 10^{-4}$ ($+$), $6.88 \times 10^{-4}$ ($\Box$). 
A line of slope $-1/4$ is shown
for comparison.}
\end{figure}

The outer geostrophic jet has finite
width $\delta$ in the limit $\epsilon \rightarrow 0$ (see expression
(\ref{dirac}) for the definition of $\epsilon$). 
Investigating the variation of $\delta$ with $\lambda$, $E$, and $E_M$
gives an useful insight into the mechanism of generation of the geostrophic
motions. Here $\delta$ is arbitrarily defined as the distance along
$s$ between the two cylinders, on both sides of the geostrophic jet,
where the angular velocity has half its
maximum value. For the range
of parameters that has been extensively explored, the outer layer is always
well characterized from $t=0.17$ onwards. Results are reported for this time.
Keeping $E$ constant and $P_m=1$, it is found that $\delta$ varies as
$\lambda ^{-1/4}$ (see figure 6). 
For fixed values of $\lambda$ and $P_m=1$, $\delta$
varies as $E_M^{1/4}$. 
This is illustrated by the figure 7. Indeed, assuming
$\delta \sim \lambda ^{-1/4}$,
the power law  $\delta \sim E_M^{1/4}$ can simply be written:
\begin{equation}
\delta \sim \left(S^\dag\right)^{-1/4}
\label{asympt}
\end{equation}
using the Lundquist number $S^\dag =\lambda (E_M)^{-1}$.
Thus, the width $\delta$ is independent of
the angular velocity $\Omega$.
We are interested by results for $P_m < 1$ since $\nu\ll\eta$ in
the geophysical case and in laboratory experiments as well.
Decreasing $P_m$ from $P_m=1$, a slight dependence of
$\delta$ on $P_m$ is found (figure 8). 
Extension of the relationship (\ref{asympt}) to solutions
for $P_m < 1$ is supported by these results.

\begin{figure}
\centerline{\hspace*{-0.05\textwidth}
\includegraphics[clip=true,width=0.8\textwidth]{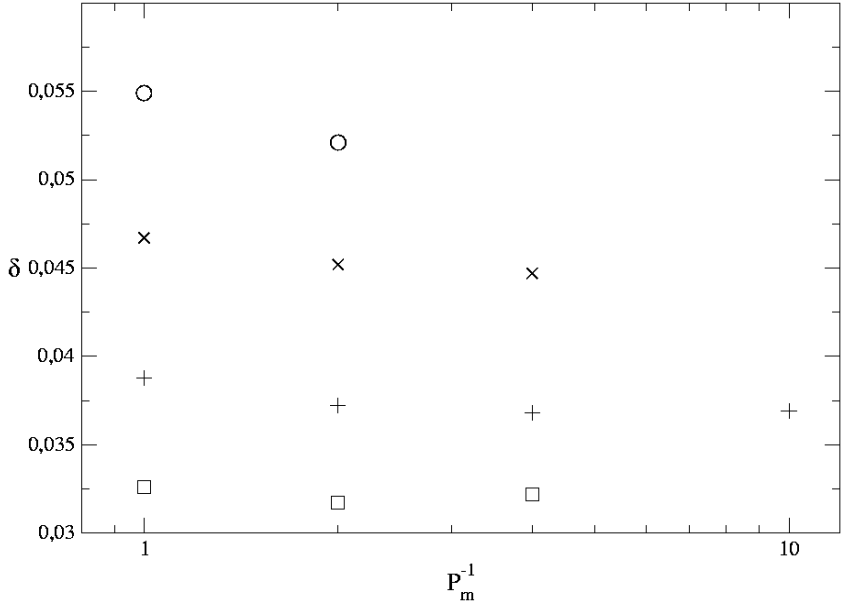}}
\caption{Thickness $\delta$ of the outer geostrophic jet with
respect to $P_m^{-1}$ for $P_m\le 1.$ and various $\lambda$: 
$\lambda=4.3 \times 10^{-5}$ (circles), $\lambda=8.6 \times 10^{-5}$ (crosses),
$\lambda=1.72 \times 10^{-4}$ (+ signs), $\lambda=3.44 \times 10^{-4}$
(squares).}
\end{figure}

The outer geostrophic jet is radiated from the tangent cylinder, which
touches the inner core on its equatorial circle. There, both the
rotation vector and the magnetic field are parallel to the inner core
surface and the
Ekman-Hartmann viscous boundary layer adjacent to the inner core is singular.
Thus, it is instructive to investigate the influence of the strength of
the magnetic field at the singularity. For this purpose, an axial uniform field
can be added to the magnetic field defined by (\ref{Bbasic}). It is found
that the jet is much thickened if the two
fields cancel out at the singularity. Then, the relationship (\ref{asympt})
does not hold.
On the other hand, it is also found - in the narrow parameter
range which has been investigated - that the
layer shrinks as $\left[B_z(b,0)\right]^{-1/4}$ as the 
axial magnetic field adjacent to the
equatorial ring of the inner core is increased. 
Taking this result at its face value,
the magnetic field strength entering the relationship (\ref{asympt}) would be
$B_z(b,0)$.
Putting these results together, it appears that the magnetic structure
adjacent to the equator of the inner core plays a significant role
in the emergence of the two propagating shear layers.
The transformation of the outer shear layer into an independent jet
detached from the inner core is promoted by the axial magnetic field.

Once the outer jet is formed, its time evolution is given by the
equations of \cite{brag70}, which satisfy
angular momentum conservation.
It is possible to estimate the angular momentum 
$ A(t)=s^2\sqrt{1-s^2}\delta u_\phi(s)$ carried
by the outer jet from the width $\delta$. For the solutions that have 
been investigated, {\it A} does not change throughout the
propagation of the jet.
The strength $B_{s}$ of the magnetic field sheared by the jet decreases with 
$s$
to 0 at $s=1$. Thus, the jet slows down as it approaches the outer sphere 
equator, which it never reachs.

The scaling (\ref{asympt}) has implications
for  the coupling between 
the axial rotation of the Earth's solid inner core and torsional oscillations
\citep{buffett05}.
The time unit $\lambda ^{-1} \Omega^{-1}$ corresponds to 1-10 years for
geophysical applications.
A characteristic time $\left(S^\dag\right)^{-1/4}(\lambda\Omega)^{-1}$ 
can be derived 
from the length $\delta$ and
the Alfv\'en wave velocity $\lambda \Omega a$. It corresponds
to the inner core rotation period below which dissipative processes at
the tangent cylinder are important. Using a geophysical estimate for
$S^\dag$, the coupling mechanism presented above between
the rotation of the inner core and torsional Alfv\'en waves is 
found to be efficient
on periods longer than a few months.
Investigation of magnetic fields with non-dipole symmetry will be a natural
follow-up of this study.

Finally, keeping $S^\dag$ constant and increasing $\lambda$, it is found 
that the structures radiated from the tangent cylinder lose their
geostrophic character for $\lambda \sim 10^{-2}$.

\section{Discussion}

Let us now examine to what extent the parameter $\lambda$
is also appropriate to comment on the geometry of fast motions in other
problems characterized by rapid rotation and magnetic field.

\subsection{Axially invariant hydromagnetic instabilities occurring 
at small Lehnert number}

We can discriminate
between two approaches that have been followed to study the stability of a 
magnetic field in a rotating sphere according to the parameter, 
either $\lambda$ or $\Lambda$, used
to compare magnetic and rotation forces. \cite{malkus67} recently followed
by \cite{zhang03} studied hydromagnetic waves in a non-dissipative
fluid ($\Lambda \rightarrow \infty$). As a result, they wrote the condition
for stability as a relationship involving the Lehnert number $\lambda$, which
has to exceed values of the order unity. \cite{zhang94} focused 
on the rapid
rotation limit ($\lambda \rightarrow 0$) instead. 
Then, the onset of instability occurs
for a critical value of the Elsasser number $\Lambda _c$. 
In accordance with the above
discussion on the geometry of the motions in the limit 
($\lambda \rightarrow 0$), they found that the (non-axisymmetric) instability
is characterized by nearly two-dimensional columnar fluid motions despite
$\Lambda _c$ being $O(10)$.

This result stands when the instabilities are thermally driven.
\cite{zhang95} focused his study of rotating convection in the presence of an
axisymmetrical toroidal magnetic field on the limit ($E \rightarrow 0$),
which amounts to ($\lambda \rightarrow 0$) for finite values of the
Elsasser number $\Lambda$. The magnetic Prandtl number is set to $1$ and
the control parameters are thus $\Lambda$ together with a Rayleigh number.
The fluid motions, in the limit ($E \rightarrow 0$), 
are almost two-dimensional 
showing only slight variations along
the direction of the rotation axis whilst
$\Lambda$ is $O(1-10)$. The solutions of \cite{olson95} (outside the 
cylindrical surface tangent to the inner core), \cite{walker97} 
(their figures 6f, 7f, 8f, 9f) for different basic states
and \cite{zhang96} (their figure 4) for $\kappa/\eta \ll 1$, where 
$\kappa$ is thermal 
diffusivity,
all present similar features.

\subsection{Columnar flow structure in geodynamo models}

Obviously, the numbers $\lambda$ and $\Lambda$ are less directly
relevant to studies of dynamo simulations than to investigations of
models with imposed large-scale magnetic field. These two estimates 
of the magnetic field strength come out as
output of the numerical runs instead of being among the initial parameters.
It is nevertheless true that
realistic values of  $\lambda$ are reached in numerical
models of the geodynamo as $\lambda$ does not involve diffusivities. 
Thus,
\cite{christensen06} conducted a statistical analysis of a set of geodynamo 
models
and estimated a parameter defined as $\lambda$, using
the shell depth as the length-scale $l$ in (\ref{lambda}). 
Their results correspond to $\lambda$  varying 
from $7\times 10^{-3}$ to $3 \times 10^{-2}$, 
keeping the core radius as length-scale.
\cite{christensen06} found that the narrow range of $\lambda$ 
values contrasts
with the wide variations of $\Lambda$.
They also remarked that measuring the relative strength
of magnetic and rotation
forces acting in geodynamo models
with the Elsasser number $\Lambda$ does not reflect the fact that 
the Lorentz force 
depends on the length scale of the magnetic field whereas the Coriolis
force is independent of the length scale of the velocity field.
Conversely, comparing typical periods of the Alfven waves to typical 
periods of the inertial waves shows that the relative importance of the 
magnetic force augments with decreasing length scales (see the expression 
(\ref{lambda}) of $\lambda$ ).
In their previous systematical parameter study, \cite{christensen99} had found
only one case with strong deviations from columnarity ($\Lambda=14$,
$E_M=6 \times 10^{-5}$, $\lambda=3 \times 10^{-2}$).
Together, these results are consistent with the statement that the extent
to which rotation affects the structure of the flow depends
on $\lambda$, the motions remaining columnar up to 
$\lambda=O(3 \times 10^{-2})$.

\subsection{Torsional oscillations in convective dynamo models}

Convection columns can excite time-dependent 
geostrophic motions in dynamo models through magnetic and
Reynolds stresses. This has been illustrated by \cite{dumberry03}.
They separated the axisymmetric azimutal velocity field
obtained from the geodynamo model of \cite{kuang99}
- $E=E_M=4 \times 10^{-5}$, $P_r=1$ and stress-free boundary conditions -
into a mean flow plus a fluctuating component.
They showed that the quasi-static azimutal winds have large gradients
in the $z$-direction. On the other hand,
\cite{dumberry03} emphasized 
the axial invariance of the time-varying zonal flows.
Their finding that, outside
the tangent cylinder, the whole
length of the geostrophic cylinders accelerates azimutally as if they
were rigid on time-scales $\tau \sim 0.1 \, \tau _D$ is in line
with the small value of $\lambda$ in this numerical experiment
($\tau _D$ magnetic diffusion time).
Indeed, using $B \sim 2 (2 \Omega \mu \rho
\eta)^{1/2}$ (see fig. 10 of \cite{kuang99}), 
we infer $\Lambda \sim 10$ and $\lambda \sim 2 \times 10^{-2}$.
These geostrophic motions
are not Alfv\'en waves as Reynolds stresses and viscous forces 
are as important as the magnetic forces in the balance of forces
acting on the geostrophic cylinders.
More recently and
with Earth-like no-slip boundary conditions, \cite{taka05} 
argued indeed that the Ekman number has to be decreased
down to $E=8 \times 10^{-6}$ to make the viscous torque acting 
on the geostrophic
cylinders negligible and the magnetic torque predominant. 
Finally, for the same value of
$E$ as \cite{taka05}, 
but with stress-free boundary conditions and small Prandtl number
$P_r=0.1$, 
\cite{busse05} found a
dynamical state where the magnetic torques on geostrophic cylinders account 
for most of the geostrophic acceleration. Extracting the average magnetic field
strength from the figure 18 of \cite{busse05} gives 
$\lambda \sim 5 \times 10^{-3}$ - well in the domain $\lambda \ll 1$ -
and $\Lambda \sim 3$. The result that
magnetic forces dominate over Reynolds stresses in the balance of force
acting on the geostrophic cylinders can be related to the observation 
that the magnetic
energy is much stronger than the kinetic energy in this solution.
Thus, sequences where geostrophic motions behave as torsional oscillations
begin to be detected in convective dynamo models. That requires $\lambda \ll 1$
- to obtain time-dependent geostrophic motions, observed for
$\lambda \sim 2 \times 10^{-2}$ by \cite{dumberry03} -, small $E$ and,
presumably,
kinetic energy weaker than magnetic energy.
In these studies, there is no evidence that the quantity $\{B_s^2\}$
measuring the intensity of the magnetic field sheared by the 
geostrophic motions
and the geostrophic velocities
evolve on separate time-scales.
Further work is needed to decide what
kind of models (fully consistent dynamo models with poor separation of scales
versus models incorporating a steady field) better describes the Earth's core 
dynamics.

\subsection{Torsional oscillations and Taylor states}

In this article, torsional oscillations are considered as part of the 
rapid motions that
are dominated by rotation because magnetic diffusion is negligible.
\cite{brag70} had a different line of arguments.
He attributed torsional oscillations to departures from
a dynamic equilibrium where the net torque of the 
Lorentz force on any geostrophic cylinder is zero \citep{taylor63}.
This condition has to be met, in spherical geometry, when only the Coriolis, 
pressure, 
buoyancy and magnetic forces are taken into account (MAC balance). 
It is frequently referred
to as a ``Taylor state''. It describes a dynamo regime on the long
time-scale for which magnetic diffusion is important.
Reinstating the acceleration of geostrophic motions
$\partial u_\phi (s)/\partial t$ in the equation
for azimutal velocities, it has been possible
to exhibit inviscid solutions of the
model-Z of \cite{brag78} that are in a Taylor state \citep{jault95}. 
However, this is almost
the unique instance where a connection between torsional oscillations
and idealized Taylor states has been vindicated.
Geodynamo numerical models showing
torsional oscillations that keep bringing back the magnetic field
towards a Taylor state have not yet been found. 
The two viewpoints differ insofar nonzonal rapid motions are
considered. I envision here that they are also constrained by rotation 
being almost $z$-invariant whereas
\cite{brag70} made no predictions on the geometry of these motions.

\section{Concluding remarks}

Focusing a numerical study on the transient motions spawned by an 
impulse in the
rotation of the inner boundary of a rapidly rotating spherical
shell immersed
in a magnetic field with dipole symmetry, I have documented 
the emergence of
geostrophic jets from the cylindrical surface that touches the
inner core at its equator, irrespectively of the value of the Elsasser number.
Both the poloidal motions, of which the vorticity is aligned
along $\mathbf{e_\phi}$, and the toroidal motions with shear
in the $z$ direction are rapidly eliminated.
The geostrophic layers travel with the velocity $\lambda a \Omega$ and
are governed, outside the tangent cylinder, 
by the equations written by \cite{brag70}. 
The jet width scales as $\left(S^\dag\right)^{-1/4}$. 
This estimate yields the frequency
below which oscillations of the solid core are communicated to
torsional Alfv\'en waves in the fluid shell. Using Earth-like parameters,
it corresponds to a period of a few months.
This study gives an illustration of the key role
played - for fast flows -
by the parameter $\lambda$ - independent of magnetic diffusivity -
put forward by \cite{lehnert54b}.
Conversely, $\lambda$ is not appropriate to study steady
solutions as it cannot be derived from the two parameters $\Lambda$ and
$E$ that characterize the static problem. 

In the same spirit, I have been able to base a discussion of
earlier numerical studies of magnetic instabilities
and dynamos in rotating shells on that parameter
$\lambda$. I suggest that
the smallness of $\lambda$  in some
of these studies is the reason for the occurrence of
columnar motions aligned parallel to the axis of rotation
and also of geostrophic flows evolving as Alfv\'en waves.
From these earlier studies, I anticipate that the results presented here
can be extended to the nonaxisymmetric case.

Thus, the parameter $\lambda$, instead of the Elsasser number $\Lambda$ ,
is the appropriate parameter to compare magnetic and rotation forces when
flows evolving on time-scales much shorter than the magnetic diffusion
time are considered.
As the value of $\lambda$ appropriate to the Earth's fluid core is 
$O(10^{-4})$,
I advocate that motions 
in the core interior
with fast diffusionless time-scales are
approximately $z$-independent and
columnar with vorticity
aligned parallel to the rotation axis. That paves the way
for dynamical studies of the flows responsible for the secular variation
of the Earth's magnetic field, generalizing to all fast motions what has
already been achieved for the geostrophic, axially symmetric ones
\citep{zatman97b}.

\section*{Acknowledgements}
I thank D. Brito, A. Fournier, H.-C. Nataf and A. Pais for their careful
reading of the initial manuscript. A. Soward and an anonymous referee are
acknowledged for their detailed reports.
This work has been supported by a grant from the French Agence Nationale de la 
Recherche, Research programme VS-QG (grant number BLAN06-2.155316).

\bibliography{pepi07}

\end{document}